\documentclass[aps,prb,floatfix,twocolumn,showpacs,preprintnumbers,amsmath,amssymb]{revtex4-1}

\usepackage{graphicx}
\usepackage{color}
\usepackage{amsmath}
\usepackage{amssymb}
\usepackage{amsthm}
\usepackage[dvipsnames]{xcolor}
\usepackage{multirow}  
\usepackage{dcolumn}   
\usepackage{bm}        
\usepackage{enumerate}
\usepackage{float}

\usepackage{hyperref}

\def  \bsig    {\mbox{\boldmath$\sigma$}}

\begin{document}

\preprint{}
\title{Thermally-induced spin polarization in a magnetized two-dimensional electron gas with
	Rashba spin-orbit interaction}
\title{Thermally-induced spin polarization in a magnetized two-dimensional electron gas with
	Rashba spin-orbit interaction}
\author{A. Dyrda\l$^{1,2}$, J.~Barna\'s$^{1,4}$, V. K.~Dugaev$^{3}$, J.Berakdar$^{2}$}
\address{$^1$Faculty of Physics, Adam Mickiewicz University,
	ul. Umultowska 85, 61-614 Pozna\'n, Poland \\ 
    $^2$ Institut f\"{u}r Physik, Martin-Luther-Universit\"{a}t Halle--Wittenberg, 06099 Halle (Saale), Germany \\
	$^3$ Department of Physics and Medical Engineering, Rzesz\'ow University of Technology, al. Powsta\'nc\'ow Warszawy 6, 35-959 Rzesz\'ow, Poland\\
	$^4$  Institute of Molecular Physics, Polish Academy of Sciences,
	ul. M. Smoluchowskiego 17, 60-179 Pozna\'n, Poland
}

\date{\today}

\begin{abstract}
Spin polarization induced by a temperature gradient (heat-current) in a magnetized two-dimensional electron gas (2DEG) with Rashba spin-orbit interaction is considered theoretically within the linear response theory. Using the Matsubara Green function formalism we calculate the temperature dependence of the spin polarization for arbitrary  orientation of the exchange field.  The limit of a nonmagnetic 2DEG (zero exchange  field) is also considered. The physical mechanisms  of the spin polarization within our scheme are  discussed.
\end{abstract}

\maketitle

\section{Introduction}

Spin-orbit interaction couples the orbital motion of an electron to its spin orientation. In conducting materials this coupling leads to various transport phenomena like anomalous Hall and Nernst effects as well as their spin counterparts, i.e. spin Hall and spin Nernst effects. These phenomena enable pure electrical or pure thermal control of spin (magnetic) moments. \cite{Sinova_RMP2015,SinovaZutic2012}. Indeed, the spin current induced by the spin Hall effect is widely functionalized as a spin torque (so called spin-Hall torque) exerted on a magnetic moment triggering a magnetic dynamics and/or  magnetic switching when the spin current exceeds a certain critical value.

One of the other consequences of the spin-orbit interaction  is the current-induced nonequilibrium spin polarization (CISP) of conduction electrons.  This means effectively that the system can be magnetically polarized by an electric field, similarly as in the case of multiferroic (magneto-electric) systems. The phenomenon of CISP  was predicted theoretically in the '70s~\cite{dyakonov71,Ivchenko78} and later it was studied theoretically~\cite{edelstein90,aronov89,liu08,gorini08,wang09,schwab10,golub11,dyrdal13,dyrdal14} and also  experimentally~\cite{vorobev1979,kato04,silov04,sih05,yang06,stern06,koehl09,kuhlen12,norman14} in various materials.
The phenomenon of CISP can occur in nonmagnetic as well as in magnetic systems, provided they exhibit spin-orbit coupling. For a magnetic system   in equilibrium the induced non-equilibrium spin polarization may couple  to the  local magnetization {\it via} the exchange interaction, and this leads to a spin torque exerted on the local magnetization~\cite{Manchon08,Abiague09,Gambardella11,Garello13,Kurebayashi}.

Recently, it has been shown that not only external electric field but also a temperature gradient may lead to the spin-orbit driven spin polarization~\cite{dyrdal13,wang10,XiaoMa2016}. These results initiated an interesting discussion on the thermally induced spin-orbit torque and also on possibly   new ways of magnetization switching as an alternative  to  switching by  electrically-induced spin transfer torque~\cite{Li04,Hatami07,Ansermet10}.
Physical mechanisms of the thermally-induced spin polarization of conduction electrons are different from that for electrically induced  spin polarization, though there are some similarities. In the case of an electric field, the field drives electrons and their wave vectors acquire a change $\Delta {\mathbf{k}}$ along the driving force. The electron spins precess to fit to new orientations of the Rashba field  creating  different distributions of electrons with positive and negative wavevector components along the electric field. Taking into account the spin precession and equilibrium spin  orientations in the two electronic subbands, one finds a nonzero net component of  the spin polarization along the in-plane axis normal to the electric field. In the case of a temperature gradient there is no electrical (mechanical) force, but instead we have a statistical force. In the absence of temperature gradient, the average spin is zero. For a finite  temperature gradient  the local distributions of electrons with positive and negative wavevector components  are different. This is because colder electrons arrive at a given point from one side and hotter electrons  from the other side. Due to different densities of states in the two Rashba subbands, a nonzero spin polarization along the axis $y$ (in-plane and normal to the gradient) appears.

Although the spin-orbit torques induced by an electric current and a temperature gradient are attracting a  great deal of attention   experimentally, for a consistent theoretical description of the spin polarization  some work is still needed.
 In this paper we consider the heat-current-induced spin polarization of a magnetic two-dimensional electron gas (2DEG) with Rashba spin-orbit interaction. Such a model is fundamental for many devices based on magnetic semiconductor heterostructures. We also briefly reconsider the nonmagnetic limit. To find the spin polarization we employ the Matsubara-Green's function formalism. Detailed numerical calculations show that the polarizability, defined as the spin polarization divided by a temperature gradient,   vanishes in the zero-temperature limit in the small impurity concentration limit. Apart from this, the spin polarization has a maximum in the range of chemical potentials where the modification of the electronic subbands are large, i.e. in the vicinity of the band edge of the lower subband.

The paper is organized as follows. In section II we describe the model and formalism used to obtain some general formulas that allow to calculate the spin polarization induced by a thermal gradient.  In section III we present and discuss the results on the spin polarization  in the absence of  exchange field. Then, in sec. IV we include the exchange field and present results for its arbitrary orientation.
Finally, in section V we summarize our results and  conclude.

\section{Model and Method} 

The two-dimensional electron gas with a spin-orbit interaction of the Rashba type and arbitrarily oriented exchange field is
described by a Hamiltonian of the form
\begin{equation}
\label{H}
H = \frac{\hbar^{2} k^{2}}{2 m} \sigma_{0} + H_{\textrm{R}} + H_{\textrm{ex}}
\end{equation}
where  $H_{\textrm{R}}$ is the Rashba term,
\begin{equation}
\label{H_R}
H_{\textrm{R}} =  \alpha (k_{y} \sigma_{x} - k_{x} \sigma_{y}),
\end{equation}
with $\alpha$ being the Rashba parameter, while $H_{\textrm{ex}}$ describes the exchange interaction
\begin{equation}
\label{H_ex}
H_{\textrm{ex}} = \mathbf{H} \cdot \bsig ,
\end{equation}
where $\mathbf{H}$ is the effective exchange field  (measured here in energy units). In the equations above the matrices  $\sigma_{0}$ and $\bsig = \{\sigma_{x}, \sigma_{y}, \sigma_{z} \}$ are the unit and Pauli matrices, respectively,  defined in the spin space. In turn,  $k_x$ and $k_y$ are the in-plane wavevector components.

We consider the  non-equilibrium spin polarization in the system driven by a statistical force, i.e. by the temperature gradient (heat current).
In this work we will consider the case of  a small, uniform temperature gradients $\mathbf{\nabla}T$ across the whole system such that the average temperature $T$ is basically constant on the scale of the magnon and carrier wavelengths. 
With these assumptions we will then employ linear response theory at finite temperatures. 
 To describe the perturbation we will resort to similar  concepts such as those introduced by Luttinger~\cite{Luttinger} and Strinati et al.~\cite{Strinati} by 
  defining   an auxiliary  time-dependent vector field of frequency $\omega/\hbar$, $\mathbf{A}(t)=\mathbf{A}(\omega)\exp (-i\omega t/\hbar) $, which is  associated with the heat current density operator, $\hat{{\bf j}}^{h} = \frac{1}{2} \left[\hat{H}-\mu \sigma_{0}, \hat{\bf v} \right]_{+}$ (here $\mu $ denotes the chemical potential), and therefore the perturbation term has the form
\begin{equation}
\hat{H}_{\mathbf{A}}^{\scriptstyle{\nabla T}}(t) = - \hat{{\bf j}}^{h} \cdot {\mathbf A}(t) .
\end{equation}
The vector field is connected to the temperature gradient {\it via}  the relation ${\mathbf A}(\omega)= \frac{\hbar}{i \omega} \left(- \frac{{\mathbf \nabla} T }{T}\right)$~\cite{dyrdal13,Ma,Tatara2015_1,Tatara2015_2,Dyrdal2016a,Dyrdal2016b}. For a  temperature gradient  along the axis $x$ the perturbation $\hat{H}_{\mathbf{A}}^{\scriptstyle{\nabla T}}(t)$ takes the form
$\hat{H}_{\mathbf{A}}^{\scriptstyle{\nabla T}}(t) = - \hat{j}^{h}_{x} A_{x}(t)$.

Within the conditions stated above, the non-equilibrium spin polarization, as a first order response  to the temperature gradient can be  calculated within the Matsubara-Green functions as
{\small{\begin{eqnarray}
\label{11}
S_{\alpha} (i \omega_{m})= \frac{1}{\beta}
\sum_{\mathbf{k}, n} \mathrm{Tr}\left\{\hat{s}_{\alpha} G_{\mathbf{k}}(i \varepsilon_{n} + i \omega_{m}) \hat{H}_{\mathbf{A}}^{\scriptstyle{\nabla T}} (i \omega_{m})G_{\mathbf{k}}(i \varepsilon_{n}) \right\}, \nonumber\\
\end{eqnarray}}}
where $\beta = 1/k_{B} T$, $\hat{s}_{\alpha}$ is the  $\alpha$-th component of the spin operator  and $\hat{H}_{\mathbf{A}}^{\scriptstyle{\nabla T}}(i \omega_{m})=-\hat{j}^{h}_{x} A_{x}(i\omega_m)$ with the amplitude of the vector potential: $A_{x}(i\omega_m)= \frac{\hbar}{i (i\omega_m)} \left(- \frac{\nabla_{x} T}{T}\right)$. Furthermore, $ \varepsilon_{n} = (2n + 1) \pi/\beta$ and $ \omega_{m} = 2 m \pi /\beta$ are the Matsubara energies, while $G_{\mathbf{k}}(i \varepsilon_{n})$ are the Matsubara-Green functions.

To sum over the Matsubara energies,  we need to assume $k_BT > \Gamma =\hbar /2\tau$, where $\Gamma$ is the imaginary part of the selfenergy, while $\tau$ is the corresponding relaxation time. Then, upon performing the summation over the Matsubara energies~\cite{abrikosov,mahan} one finds the  spin polarization induced by the temperature gradient in the following form:
\begin{eqnarray}
\label{S_alpha}
S_{\alpha} (\omega)= \hspace{7.2cm}\nonumber \\
\frac{\hbar}{\omega} \frac{\nabla_{x} T}{T} {\mathrm{Tr}} \sum_{\mathbf{k}} \int \frac{d \varepsilon}{2 \pi} f(\varepsilon) \hat{s}_{\alpha} \Bigl(  G_{\mathbf{k}}^{R}(\varepsilon + \omega) \hat{j}_{x}^{h} [G_{\mathbf{k}}^{R}(\varepsilon) - G_{\mathbf{k}}^{A}(\varepsilon)] \Bigr. \nonumber\\
+ \Bigl.  [G_{\mathbf{k}}^{R}(\varepsilon) - G_{\mathbf{k}}^{A}(\varepsilon)] \hat{j}_{x}^{h}G_{\mathbf{k}}^{A}(\varepsilon - \omega) \Bigr).\hspace{0.5cm}
\end{eqnarray}
The key steps of the derivation of the above formula are described elsewhere\cite{Dyrdal2016a,Dyrdal2017}. Equation (\ref{S_alpha}) is our starting expression for further considerations.

Since the spin polarization is linear in $\nabla_x T$, we may also define the thermal spin polarizability as
\begin{equation}
{\cal{P}} (\omega) =S_y (\omega)/\nabla_x T,
\end{equation}
so the polarizability can be calculated from the formula
\begin{eqnarray}
\label{polar}
{\cal{P}} (\omega)= \hspace{7.2cm}\nonumber \\
\frac{\hbar}{\omega} \frac{1}{T} {\mathrm{Tr}} \sum_{\mathbf{k}} \int \frac{d \varepsilon}{2 \pi} f(\varepsilon) \hat{s}_{\alpha} \Bigl(  G_{\mathbf{k}}^{R}(\varepsilon + \omega) \hat{j}_{x}^{h} [G_{\mathbf{k}}^{R}(\varepsilon) - G_{\mathbf{k}}^{A}(\varepsilon)] \Bigr. \nonumber\\
+ \Bigl.  [G_{\mathbf{k}}^{R}(\varepsilon) - G_{\mathbf{k}}^{A}(\varepsilon)] \hat{j}_{x}^{h}G_{\mathbf{k}}^{A}(\varepsilon - \omega) \Bigr).\hspace{0.5cm}
\end{eqnarray}

\section{Spin polarization in a nonmagnetic 2DEG}

Let us consider first the case with zero exchange field, i.e., when the 2DEG is nonmagnetic.  The Hamiltonian (\ref{H})  reduces  to the following form
\begin{equation}
\label{HR}
H = \frac{\hbar^{2} k^{2}}{2 m} \sigma_{0} + \alpha (k_{y} \sigma_{x} - k_{x} \sigma_{y}),
\end{equation}
while the corresponding impurity-averaged retarded/advanced ($R/A$) Green function can be written in the form
\begin{eqnarray}
\label{Gf_HR}
G_{\mathbf{k}}^{R/A}(\varepsilon) = G_{\mathbf{k}\, 0}^{R/A}(\varepsilon) \sigma_{0} + G_{\mathbf{k}\, x}^{R/A}(\varepsilon) \sigma_{x} + G_{\mathbf{k}\, y}^{R/A} (\varepsilon) \sigma_{y},\hspace{0.4cm}
\end{eqnarray}
where
\begin{subequations}
	\begin{align}
	G_{\mathbf{k}\, 0}^{R/A}(\varepsilon) &= \frac{1}{2} [G_{+}(\varepsilon) + G_{-}(\varepsilon)], \\
	G_{\mathbf{k}\, x}^{R/A}(\varepsilon) &= \frac{\alpha k_{y}}{2 \lambda_{k}}[G_{+}(\varepsilon) -G_{-}(\varepsilon)],\\
	G_{\mathbf{k}\, y}^{R/A}(\varepsilon) &= - \frac{\alpha k_{x}}{2 \lambda_{k}} [G_{+}(\varepsilon)-G_{-}(\varepsilon)],
	\end{align}
\end{subequations}
with $G_{\pm}^{R}(\varepsilon) = [\varepsilon + \mu - E_{\pm}  + i\Gamma]^{-1}$, $G_{\pm}^{A}(\varepsilon) = [\varepsilon + \mu - E_{\pm}  - i\Gamma]^{-1}$ and $E_{\pm} = \frac{h^{2} k^{2}}{2m} \pm \lambda_{k}$ (with $\lambda_{k} = \alpha k)$. Note, the relaxation rate $\Gamma$  in a nonmagnetic electron gas with Rashba interaction (assuming relaxation due to  scattering on short-range impurities only) is constant for $\mu >0$ and energy dependent for $\mu <0$. In this paper, however, we assume $\Gamma$ as a constant parameter.

\subsection{Bare bubble  approximation}

The heat current operator corresponding to the Hamiltonian (\ref{HR}) has the explicit form
\begin{eqnarray}
\label{jh_HR}
\hat{j}_{x}^{h} = \Bigl( (\varepsilon_{k} - \mu) \frac{\hbar k_{z}}{m} +  \frac{\alpha^2}{\hbar} k_{x} \Bigr) \sigma_{0}\hspace{2cm}\nonumber\\
+ \alpha \frac{\hbar}{m} k_{x} k_{y} \sigma_{x}
- \Bigl( \frac{\alpha}{\hbar} (\varepsilon_{k} - \mu) + \alpha \frac{\hbar}{m} k_{x}^{2} \Bigr) \sigma_{y}.
\end{eqnarray}
Inserting Eqs. (\ref{Gf_HR}) to (\ref{jh_HR}) into Eq.(\ref{S_alpha}) we find that only the $y$ component of the spin polarization is non-zero, namely 
\begin{eqnarray}
S_{y}^{T}(\omega) = \frac{\hbar}{\omega} \frac{\nabla_{x}T}{T} \int \frac{dk}{(2\pi)^{2}} \left( -\frac{\pi}{2} [2 \varepsilon_{k}(\varepsilon_{k} - \mu) + \alpha^{2} k^{2}]\mathcal{S}_{A}\right.\nonumber\\
\left. + \alpha \frac{\pi}{2} k [(\varepsilon_{k} - \mu) \mathcal{S}_{B} + (3 \varepsilon_{k} - \mu) \mathcal{S}_{C}] \right),\hspace{1cm}
\end{eqnarray}
where
\begin{eqnarray}
\mathcal{S}_{A} = I_{--}^{RA}(\omega) - I_{--}^{RR}(\omega) + I_{++}^{RR}(\omega) - I_{++}^{RA}(\omega)\hspace{0.9cm}\nonumber\\
+ I_{--}^{AA}(-\omega) - I_{++}^{AA}(-\omega) - I_{--}^{RA}(-\omega) + I_{++}^{RA}(-\omega),\hspace{0.2cm}
\end{eqnarray}
\begin{eqnarray}
\mathcal{S}_{B} = I_{-+}^{RA}(\omega) - I_{-+}^{RR}(\omega) + I_{+-}^{RA}(\omega) - I_{+-}^{RR}(\omega)\hspace{0.9cm}\nonumber\\
+ I_{-+}^{AA}(-\omega) + I_{+-}^{AA}(-\omega) - I_{-+}^{RA}(-\omega) - I_{+-}^{RA}(-\omega), \hspace{0.2cm}
\end{eqnarray}
\begin{eqnarray}
\mathcal{S}_{C} = I_{--}^{RA}(\omega) - I_{--}^{RR}(\omega) + I_{++}^{RA}(\omega) - I_{++}^{RR}(\omega)\hspace{0.9cm}\nonumber\\
+ I_{--}^{AA}(-\omega) + I_{++}^{AA}(-\omega) - I_{--}^{RA}(-\omega) - I_{++}^{RA}(-\omega). \hspace{0.2cm}
\end{eqnarray}
Here we use the notation $I_{mn}^{XY}(\omega) = \int \frac{d\varepsilon}{2\pi}G_{m}^{X}(\varepsilon + \omega)G_{n}^{Y}(\varepsilon)$ and $I_{mn}^{XY}(-\omega) = \int \frac{d\varepsilon}{2\pi}G_{m}^{X}(\varepsilon)G_{n}^{Y}(\varepsilon-\omega)$ with  $m,n =$ $ \{+,-\}$ and  $X,Y = \{R,A\}$.\\

Upon integrating over $\varepsilon$ and taking the limit $\omega \rightarrow 0$ we  find the thermally-induced spin polarization in the following form:
\begin{eqnarray}
\label{Sy_HR_num}
S_{y} = - \frac{\hbar}{2 \Gamma} \frac{\nabla_{x}T}{T} \int \frac{dk}{4\pi} \left(\varepsilon_{k} (\varepsilon_{k} - \mu) + \frac{1}{2} \alpha^{2} k^{2}\right)\hspace{0.9cm}\nonumber\\ \times[f'(E_{+}) - f'(E_{-})]\hspace{4cm} \nonumber\\
- \frac{\hbar}{2 \Gamma} \frac{\nabla_{x}T}{T} \frac{\alpha}{2} \int \frac{dk}{4\pi} k (3\varepsilon_{k} - \mu) [f'(E_{+}) + f'(E_{-})]\hspace{0.3cm}\nonumber \\
- \alpha \hbar \Gamma \frac{\nabla_{x}T}{T} \int \frac{dk}{4\pi} k (\varepsilon_{k} - \mu) \frac{f'(E_{+}) + f'(E_{-})}{(E_{+} - E_{-})^{2} + (2\Gamma)^{2}}.\hspace{0.3cm}
\end{eqnarray}
The first two terms in Eq.(17) are proportional to $\hbar /2\Gamma =\tau$, while the third term is proportional to $\Gamma$ (or $1/\tau$). Thus, one may expect that the first two terms are dominant in  general, while the third term is small.
 This however, does not hold true in the low temperature regime, where the first two terms cancel each other so the dominant (though very small)  contribution stems from the third term. 
 This contribution however is canceled by the impurity vertex corrections as will be shown in the following. The formula (17) is our general result for the spin polarization in the bare bubble approximation. Note, $\Gamma$ is here a parameter which is constant (independent of energy/wavevector).

In the low temperature limit one can replace the derivatives of the Fermi distribution functions by appropriate Dirac delta-functions, and then the above expression can be integrated analytically. As already mentioned, the only contribution originates  then from the third term in Eq.(17). Assuming $\mu > 0$ and taking into account the fact that the Dirac delta-functions for $\mu > 0$  can be expressed as
\begin{equation}
\delta(E_{\pm} - \mu) = \frac{m\, \delta(k - k_{\pm})}{\sqrt{2 m \mu \hbar^{2} + m^{2} \alpha^{2}} },
	\end{equation}
we find
\begin{eqnarray}\label{SyT0_f}
S_{y}
=  \frac{\nabla_{x}T}{T} \frac{1}{2\Gamma}  \frac{\alpha \hbar^{3}}{16 \pi \sqrt{2 m \mu \hbar^{2} + \alpha^{2} m^{2}}}\hspace{3cm}
\nonumber\\
\times \left[\frac{k_{+}^{3} - \frac{2 m \mu}{\hbar^{2}} k_{+}}{1 + (\alpha k_{+}/\Gamma)^{2}} + \frac{k_{-}^{3} - \frac{2 m \mu}{\hbar^{2}} k_{-}}{1 + (\alpha k_{-}/\Gamma)^{2}}\right].\hspace{0.5cm}
\end{eqnarray}
Since the formalism assumes well defined quasiparticles, the above formula is applicable for $\Gamma \ll \alpha k_{\pm}$, where $k_{\pm}=\mp \frac{m\alpha}{\hbar^2}+\frac{1}{\hbar^2}\sqrt{m^2\alpha^2 +2m\mu\hbar^2}$ are the Fermi wavevectors in the two subbands.
Taking additionally into account that $\Gamma < k_BT$,
one finds from the above equation a small, though nonzero,  spin polarization in the limit of $T\to 0$.
This holds true for  $\Gamma \ll \alpha k_{\pm}$ and $\Gamma < k_BT$.

When only one subband is occupied, $\mu < 0$, the Dirac Delta-functions for the $E_{-}$ band read
\begin{equation}
\delta(E_{-} - \mu) = \frac{m \left[\delta(k - k_{-}^{+}) + \delta(k - k_{-}^{-})\right]}{\sqrt{2 m \mu \hbar^{2} + m^{2} \alpha^{2}}},
\end{equation}
where now $k_{-}^{\pm} = \frac{m \alpha}{\hbar^{2}} \pm  \sqrt{2 m \mu \hbar^{2} + m^{2} \alpha^{2}}$.
The spin polarization is then given by the formula
\begin{eqnarray}\label{SyT0_f_muneg}
S_{y}
=  \frac{\nabla_{x}T}{T} \frac{1}{2\Gamma}  \frac{\alpha \hbar^{3}}{16 \pi \sqrt{2 m \mu \hbar^{2} + \alpha^{2} m^{2}}}\hspace{3cm}
\nonumber\\
\times \left[\frac{(k_{-}^{+})^{3} - \frac{2 m \mu}{\hbar^{2}} k_{-}^{+}}{1 + (\alpha k_{-}^{+}/\Gamma)^{2}} + \frac{(k_{-}^{-})^{3} - \frac{2 m \mu}{\hbar^{2}} k_{-}^{-}}{1 + (\alpha k_{-}^{-}/\Gamma)^{2}}\right],\hspace{0.5cm}
\end{eqnarray}
and  may generally remain small but finite in the zero-temperature limit for $\Gamma < k_BT$ and  $\Gamma << \alpha k_{-}^{\pm}$.
However, the spin polarization given by Eq.(19) as well as by Eq.(21) is canceled by the vertex corrections, as will be shown below, so spin polarization vanishes in the zero temperature limit.

\subsection{Vertex correction}

It is known that impurity vertex corrections can have a significant impact on various physical quantities, like for instance on the spin Hall conductivity of 2DEG with Rashba interaction. Therefore, we consider now the vertex corrections to the spin polarization.

The equation for the renormalized spin vertex  reads 
\begin{equation}
\label{VertexEq}
\bar{S}_{y} = \frac{\hbar}{2} \sigma_{y} + n_{i} v_{0}^{2} \int \frac{d^{2} \mathbf{k}}{(2\pi)^{2}} G_{\mathbf{k}}^{A}(\varepsilon) \bar{S}_{y} G^{R}_{\mathbf{k}}(\varepsilon + \omega).
\end{equation}
We look for the solution of Eq.(\ref{VertexEq}) in the following form: $\bar{S}_{y} = a \sigma_{0} + b \sigma_{x} + c\sigma_{y} + d \sigma_{z}$. Thus, we find that: $a=b=d=0$ while $c$ is given by the following formula:
\begin{equation}
c = \frac{\hbar}{2} \left[ 1 - \frac{1}{2} \pi n_{i} V^{2}\left( \mathcal{I}_{--}^{RA} + \mathcal{I}_{++}^{RA} + \mathcal{I}_{+-}^{RA} + \mathcal{I}_{-+}^{RA}\right)\right]^{-1}.\hspace{0.4cm}
\end{equation}
Taking the above expression at the Fermi level ($\varepsilon = 0$) and assuming the limit of $\omega \rightarrow 0$, one finds the  solution
\begin{equation}
\bar{S}_{y} = \frac{\hbar}{2} \frac{1}{1 - \frac{1}{2} \pi n_{i} v_{0}^{2} (\mathcal{I}_{1} + \mathcal{I}_{2})}\sigma_y ,
\end{equation}
where the integrals $\mathcal{I}_{1,2}$ are introduced as 
\begin{eqnarray}
\mathcal{I}_{1} = \int\frac{dkk}{(2\pi)^{2}} \left[ \frac{1}{(\mu - E_{+})^{2} + \Gamma^{2}} +  \frac{1}{(\mu - E_{-})^{2} + \Gamma^{2}} \right],\hspace{0.5cm}
\end{eqnarray}
\begin{eqnarray}
\mathcal{I}_{2} = \Re \int \frac{dkk}{(2\pi)^{2}} \frac{E_{-} - E_{+} - 2 i \Gamma}{(E_{-} - E_{+})^{2} + (2 \Gamma)^{2}}\hspace{1.5cm}\nonumber\\ \times \left[ \frac{\mu - E_{-} + i \Gamma}{(\mu - E_{-})^{2} + \Gamma^{2}} - \frac{\mu - E_{+} - i \Gamma}{(\mu - E_{+})^{2} + \Gamma^{2}}\right]\nonumber\\
+ \int \frac{dkk}{(2\pi)^{2}} \frac{E_{+} - E_{-} - 2 i \Gamma}{(E_{+} - E_{-})^{2} + (2 \Gamma)^{2}}\hspace{1.5cm}\nonumber\\ \times \left[ \frac{\mu - E_{+} + i \Gamma}{(\mu - E_{+})^{2} + \Gamma^{2}} - \frac{\mu - E_{-} - i \Gamma}{(\mu - E_{-})^{2} + \Gamma^{2}}\right].
\end{eqnarray}
It is convenient to introduce the parameter $\beta$ by the following equality:
\begin{eqnarray}
\frac{1}{2} \pi n_{i} v_{0}^{2} (\mathcal{I}_{1} + \mathcal{I}_{2})
\equiv \frac{1}{2} + \beta , \hspace{0.3cm}
\end{eqnarray}
which can be determined from Eqs.(25) and (26).
Accordingly, we may write the renormalized vertex function as
\begin{equation}
\label{A10}
\bar{S}_{y} = \hat{s}_{y} \frac{2}{1 - 2\beta},
\end{equation}
or alternatively
\begin{equation}
\label{sy_ren_op}
\bar{S}_{y} =  \hat{s}_{y} + \delta \hat{s}_{y}  \equiv \hat{s}_{y} + \gamma \frac{\hbar}{2} \sigma_{y},
\end{equation}	
where $\gamma = \frac{1 + 2 \beta}{1 - 2 \beta}$.
Thus, the spin polarization with the vertex correction included can be written in the following form:
\begin{equation}
S_{y}^{\textrm{tot}} = S_{y} + \Delta S_{y},
\end{equation}	
where $S_{y}$ is given by Eq.(\ref{Sy_HR_num}) and $\Delta S_{y}$ is  defined as:
\begin{eqnarray}
\Delta S_{y} = - \frac{\hbar}{\omega} \frac{\nabla_{x} T}{T} \textrm{Tr} \int \frac{d^{2}\mathbf{k}}{(2\pi)^{2}} \int \frac{d \varepsilon}{2\pi} f(\varepsilon) \delta\hat{s}_{y} \hspace{1.2cm}\nonumber\\
\times \left( G_{\mathbf{k}}^{R}(\varepsilon + \omega) \hat{j}_{x}^{h} G_{\mathbf{k}}^{A}(\varepsilon) - G_{\mathbf{k}}^{R}(\varepsilon) \hat{j}_{x}^{h} G_{\mathbf{k}}^{A}(\varepsilon - \omega) \right).\hspace{0.3cm}
\end{eqnarray}
Upon integrationg over $\varepsilon$ in Eq.(31), the final expression for the spin polarization is cast as 
\begin{eqnarray}
\label{Sy_HR_num_Ren}
S_{y} = - \frac{\hbar}{2 \Gamma} \frac{\nabla_{x}T}{T} (1 + \gamma) \int \frac{dk}{4\pi} \left(\varepsilon_{k} (\varepsilon_{k} - \mu) + \frac{1}{2} \alpha^{2} k^{2}\right)\hspace{0.9cm}\nonumber\\ \times[f'(E_{+}) - f'(E_{-})]\hspace{1.0cm} \nonumber\\
- \frac{\hbar}{2 \Gamma} \frac{\nabla_{x}T}{T} (1 + \gamma) \frac{\alpha}{2} \int \frac{dk}{4\pi} k (3\varepsilon_{k} - \mu) [f'(E_{+}) + f'(E_{-})]\hspace{1.0cm}\nonumber\\
- \alpha \hbar \Gamma \frac{\nabla_{x}T}{T} (1 - \gamma) \int \frac{dk}{4\pi} k (\varepsilon_{k} - \mu) \frac{f'(E_{+}) + f'(E_{-})}{(E_{+} - E_{-})^{2} + (2\Gamma)^{2}}.\hspace{0.7cm}
\end{eqnarray}
Thus, the vertex renormalization of the first and second terms in Eq.(17) for the spin polarization in the bare bubble approximation leads to their multiplication by a factor $1+\gamma = 2/(1 - 2\beta)$. This renormalization factor is equal to that found in the case of the spin polarization induced by an external electric field~\cite{Dyrdal2017}.
Note, the parameter $\beta$ in Eqs.~(27) and (28)  is equivalent  to the parameter $\beta$ introduced in   Ref.\onlinecite{Dyrdal2017}. It is  also worth  noting that in Ref.\onlinecite{Dyrdal2017} the velocity vertex was renormalized, whereas here we renormalized the spin vertex. In turn, the third term in Eq.(17) is renormalized by the factor $1-\gamma$, meaning that  it is significantly reduced in a general  and canceled in the zero temperature limit, as shown below.

In the low temperature limit the integrals (25) and (26) have the form 
\begin{eqnarray}
\mathcal{I}_{1}
\cong \frac{1}{2 \Gamma} \int \frac{dk k}{2 \pi} \left[ \delta(E_{+} - \mu) + \delta(E_{-} - \mu) \right],
\end{eqnarray}
\begin{eqnarray}
\mathcal{I}_{2}
\cong\int \frac{dk k}{2 \pi} \frac{2 \Gamma}{(2\alpha k)^{2} + (2 \Gamma)^{2}} \left[ \delta(E_{+} - \mu) + \delta(E_{-} - \mu) \right]. \hspace{0.5cm}
\end{eqnarray}
The parameter $\beta$ is then given by the formula
\begin{equation}
\beta = \frac{\hbar^{2}}{4 \sqrt{2 m \mu \hbar^{2} + m^{2} \alpha^{2}}} \left[ \frac{k_{+}}{1 + (\frac{\alpha k_{+}}{\Gamma})^{2}} + \frac{k_{-}}{1 + (\frac{\alpha k_{-}}{\Gamma})^{2}} \right]\hspace{0.5cm}
\end{equation}
for $\mu > 0$, and
\begin{equation}
\beta = \frac{\hbar^{2}}{4 \sqrt{2 m \mu \hbar^{2} + m^{2} \alpha^{2}}} \left[ \frac{k_{-}^{+}}{1 + (\frac{\alpha k_{-}^{+}}{\Gamma})^{2}} + \frac{k_{-}^{-}}{1 + (\frac{\alpha k_{-}^{-}}{\Gamma})^{2}} \right]\hspace{0.5cm}
\end{equation}
for $\mu < 0$.
The spin polarization is then given by Eq.(19) or Eq.(22), with the prefactor $1-\gamma$. This prefactor vanishes in the  zero
temperature limit for $\Gamma <k_BT$ and $\Gamma << \alpha k_{\pm}$ (or $\Gamma << \alpha k_{-}^\pm$), and so does also the spin polarization,
\begin{equation}
S_y=0
\end{equation}
for $T=0$.

\subsection{Numerical results}

The numerical results presented here are for low impurity case where the vertex corrections are irrelevant and the spin polarization is described  relatively well by the bare bubble approximation. 
All qualitative features of the spin polarization remain valid also when the vertex corrections are relevant.

\begin{figure}[t]
 	\includegraphics[width=\columnwidth]{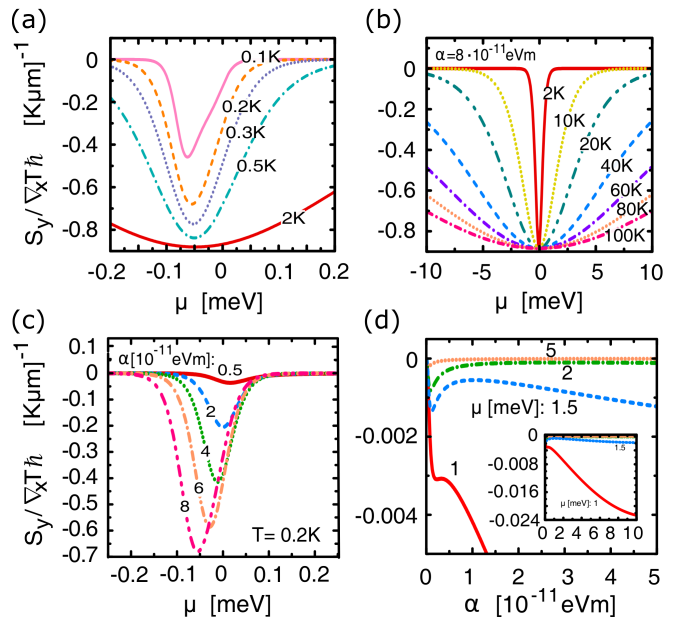}
 	\caption{The y-component of the thermally-induced   spin polarization in a nonmagnetic electron gas shown as a function of the chemical potential for the indicated temperatures (a,b); as a function of the chemical potential for the  indicated Rashba parameters (c), and as a function of the Rashba parameter for the indicated chemical potentials (d). Other parameters as indicated, whereas $m = 0.07m_{0}$, and $\Gamma =5 \cdot 10^{-3}$ meV. }
 	\label{fig:fig1}
\end{figure}

In Fig.\ref{fig:fig1} we show  the spin polarization  induced by a temperature gradient in a nonmagnetic system.
The spin polarization is normalized there to $\hbar\nabla T$, so effectively these figures show the thermal spin polarizability of the system.
The only nonzero component in the absence of the exchange field is the in-plane component perpendicular to $\nabla T$, i.e. the component $S_y$. Figures \ref{fig:fig1}a  and \ref{fig:fig1}b show the spin polarization as a function of chemical potential for different temperatures -- from very low up to 100K.  Note, the parameter $\Gamma$ assumed in Fig.\ref{fig:fig1} is 0.005 meV (corresponding to approximately 0.05 K). The lowest temperature in Fig.1 is 0.1K, i.e. the thermal energy is above $\Gamma$ for all curves,  $\Gamma <k_BT$.
When $T$ increases, the spin polarization also increases and has a maximum for the Fermi level around the bottom of the lower electronic band. The spin polarization as a function of the chemical potential has then the form of a narrow and asymmetric peak. When $T$ increases further, the maximum value of the spin polarization saturates, while  the peaks become broader. This behaviour is consistent with the physical mechanism of the thermally-induced spin polarization. Three ingredients of this mechanism are important: (i) spin orientation in the two electronic subbands is determined by the Rashba coupling and in total spin in equalibrium vanishes in each subband; (ii) Rashba splitting of the electronic bands introduces some asymmetry in the density of states of the two subbands; (iii) due to the temperature gradient, there is an imbalance in the  spin flowing into a certain region from the colder and hotter sides. All this leads to a net spin polarization. Moreover, this also explains why the spin polarization vanishes at $T\to 0$ and why its maximum appears close to the bottom of the lower band. The latter takes place because modifications in the electronic structure by the Rashba coupling are most significant there. In turn, broadening of the peaks with increasing  temperature is a consequence of the broadening of the Fermi distribution function. The almost symmetrical shape is due to the assumption of a constant chemical potential. In other words, our system is assumed to be attached to two (left and right) electronic reservoirs where electrons are described by the chemical potential $\mu$. Thus, even if the chemical potential is below the band edge in 2DEG, electrons can be injected into the system from the reservoirs when the temperature is sufficiently high.

Increasing the Rashba parameter the spin polarization also increases. Moreover, the band edge of the lower band is shifted down so the peak in the spin polarization slightly shifts towards lower values of the chemical potential. This behavior is shown in Fig.~\ref{fig:fig1}c, where different curves correspond to different values of the parameter $\alpha$. The interplay of this shift and the increase of the maximum spin polarization with $\alpha$ lead to some non-monotonous behavior of the  spin polarization with the Rashba parameter, especially at higher values of $\mu$, where the spin polarization is already very small. This is shown in Fig.~\ref{fig:fig1}d, where the normalized spin polarization is presented as a function of the Rashba parameter for the indicated values of the chemical potential. Obviously, all curves start at $S_y=0$ as the spin polarization vanishes in the absence of the Rashba coupling. In general, these curves reflect the behavior of the spin polarization shown in Figs~\ref{fig:fig1}a,b and Fig.~\ref{fig:fig1}c. Since the temperature is relatively low in Fig~\ref{fig:fig1}d and the spin polarization has a maximum around the bottom edge of the lower band,  the largest spin polarization  appears for the lowest value of the chemical potential (the red curve in Fig.\ref{fig:fig2}d). The drop in the spin polarization after the initial increase with increasing the Rashba parameter is a consequence of the decrease in the energy of the lower band edge and the corresponding shift  of the maximum in the spin polarization towards the lower chemical potentials. Note, the spin polarization in Fig.\ref{fig:fig2}d is very small.

\section{Spin polarization in the presence of exchange field}
Considering the full Hamiltonian (\ref{H}) with the exchange term $H_{\textrm{ex}}$, we cast  the exchange field in spherical coordinates
 $\mathbf{H} = (H_{x}, H_{y}, H_{z})$  as
\begin{subequations}
	\begin{align}
	H_{x} & = J M \sin\theta \cos\xi ,\\
	H_{y} & = J M \sin\theta \sin\xi ,\\
	H_{z} & = J M \cos\theta ,
	\end{align}
\end{subequations}
where  $J$ is a parameter proportional to the exchange constant and $M$ is effective magnetization that in general depends on temperature according to   Bloch's law, $M(T) = M_{0}[1 - (T/T_{c})^{3/2}]$ (with $T_{c}$ denoting the Curie temperature and $M_{0}$ standing for the magnetization at $T = 0$). The angles $\theta$ and $\xi$ are the polar and azimuthal angles in the spherical coordinates.

The eigenvalues of the Hamiltonian (\ref{H}) take now the form:
\begin{eqnarray}
E_{\pm}
= \varepsilon_{k} \pm \lambda_{\mathbf{k}} ,
\end{eqnarray}
where $\varepsilon_{k} = \frac{\hbar^{2} k^{2}}{2 m}$ (with $k^{2} = k_{x}^{2} + k_{y}^{2}$) and $\lambda_{\mathbf{k}}$ is defined as $\lambda_{\mathbf{k}} = \sqrt{J^{2} M^{2} + \alpha^{2} k^{2} + 2 J M \alpha \sin\theta (k_{x} \sin\xi - k_{y} \cos\xi)}$.
The retarded/advanced Green function corresponding to the Hamiltonian (1)  takes the following explicit form:
\begin{eqnarray}
\label{Gf_H}
G_{\mathbf{k}}^{R/A}(\varepsilon) = G_{\mathbf{k}\, 0}^{R/A}(\varepsilon) \sigma_{0}\hspace{4.5cm}\nonumber\\ + G_{\mathbf{k}\, x}^{R/A}(\varepsilon) \sigma_{x} + G_{\mathbf{k}\, y}^{R/A} (\varepsilon) \sigma_{y} + G_{\mathbf{k}\, z}^{R/A}(\varepsilon) \sigma_{z},\hspace{0.4cm}
\end{eqnarray}
where
\begin{subequations}
	\begin{align}
	G_{\mathbf{k}\, 0}^{R/A}(\varepsilon) &= \frac{1}{2} [G_{+}(\varepsilon) + G_{-}(\varepsilon)], \\
	G_{\mathbf{k}\, x}^{R/A}(\varepsilon) &= \frac{1}{2 \lambda_{\mathbf{k}}} (\alpha k_{y} + H \sin\theta \cos\xi)[G_{+}(\varepsilon) -G_{-}(\varepsilon)],\\
	G_{\mathbf{k}\, y}^{R/A}(\varepsilon) &= - \frac{1}{2 \lambda_{\mathbf{k}}} (\alpha k_{x} - H \sin\theta \sin\xi)[G_{+}(\varepsilon)-G_{-}(\varepsilon)],\\
	G_{\mathbf{k}\, z}^{R/A}(\varepsilon) &=  \frac{1}{2 \lambda_{\mathbf{k}}} H \cos\theta [G_{+}(\varepsilon) -G_{-}(\varepsilon)],
	\end{align}
\end{subequations}
while $G_{\pm}^{R}(\varepsilon) = [\varepsilon + \mu - E_{\pm}  + i\Gamma]^{-1}$ and $G_{\pm}^{A}(\varepsilon) = [\varepsilon + \mu - E_{\pm}  - i\Gamma]^{-1}$. Note,  we assumed here  $\Gamma$ as a constant parameter for both subbands.
Apart from this, the expansion of the Green function in Pauli materices includes now the term proportional to $\sigma_{z}$, which was absent in the case of no exchange field, see Eq.(8).

The
operator of the heat current density has the following explicit form:
\begin{equation}
\label{jh_H}
\hat{j}^{h}_{x} = 	j_{x0}^{h} \sigma_{0} + j_{xx}^{h} \sigma_{x} + j_{xy}^{h} \sigma_{y} + j_{xz}^{h} \sigma_{z} ,
\end{equation}
where
\begin{subequations}
	\begin{align}
	j_{x0}^{h} &= \Bigl( (\varepsilon_{k} - \mu) \frac{\hbar k_{x}}{m} +  \frac{\alpha}{\hbar}(\alpha k_{x} -  H_{y}) \Bigr),\hspace{0.1cm}\\
	j_{xx}^{h} &=  \frac{\hbar}{m} k_{x} (\alpha k_{y} + H_{x}).\hspace{2.4cm} \\
	j_{xy}^{h} &= \Bigl( \frac{\alpha}{\hbar} (\varepsilon_{k} - \mu) + \frac{\hbar}{m} k_{x} (\alpha k_{x} - H_{y})\Bigr),\\
	j_{xz}^{h} &=  -\frac{\hbar}{m} k_{x} H_{z} \sigma_{z} .\hspace{3.25cm}
	\end{align}
\end{subequations}

\subsection{General formula for the components of spin polarization}

 \begin{figure*}[t]
	\includegraphics[width=0.95\textwidth]{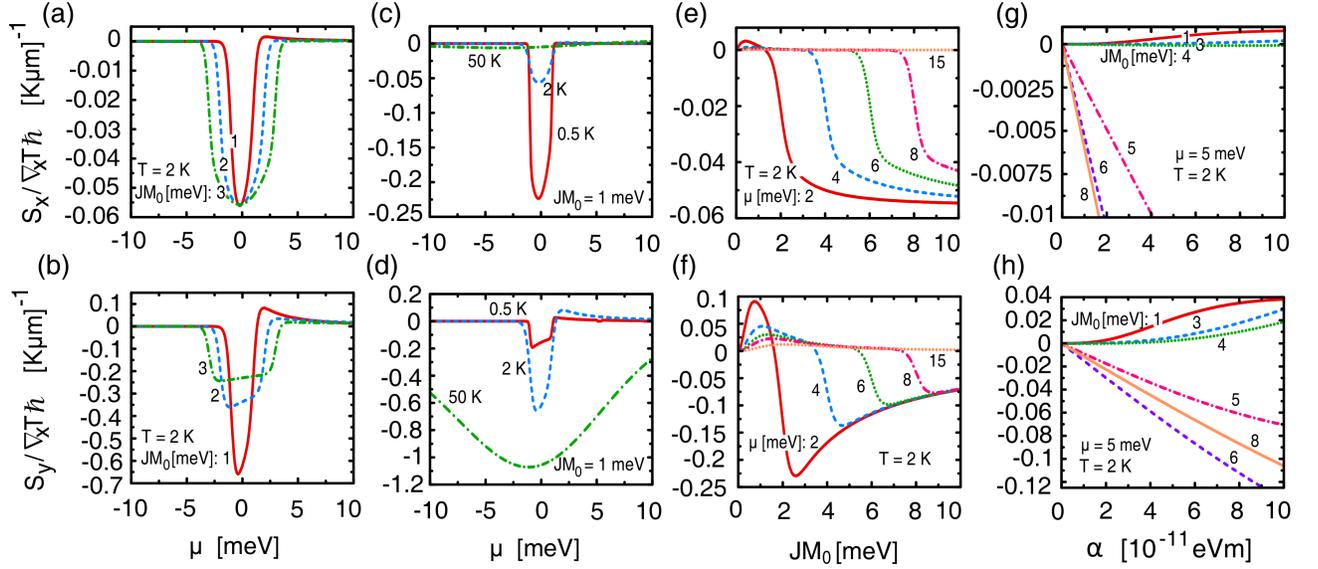}
	\caption{The $x$- and $y$-components of the spin polarization in a magnetic 2DEG. The upper panel (a,c,e,g) corresponds to the $S_x$ component while the lower one (b,d,f,h) to the $S_y$ component. Dependence on the chemical potential (a,b,c,d), magnitude of exchange field $JM$ (e,f), and Rashba parameter $\alpha$ (g,h) is shown for indicated  parameters, and for $\Gamma =5 \cdot 10^{-3}$ meV, $m=0.07 m_0$,  and $\alpha = 8 \cdot 10^{-11}$ eVm  (if not indicated otherwise). }
	\label{fig:fig2}
\end{figure*}

 \begin{figure}[t]
	\includegraphics[width=\columnwidth]{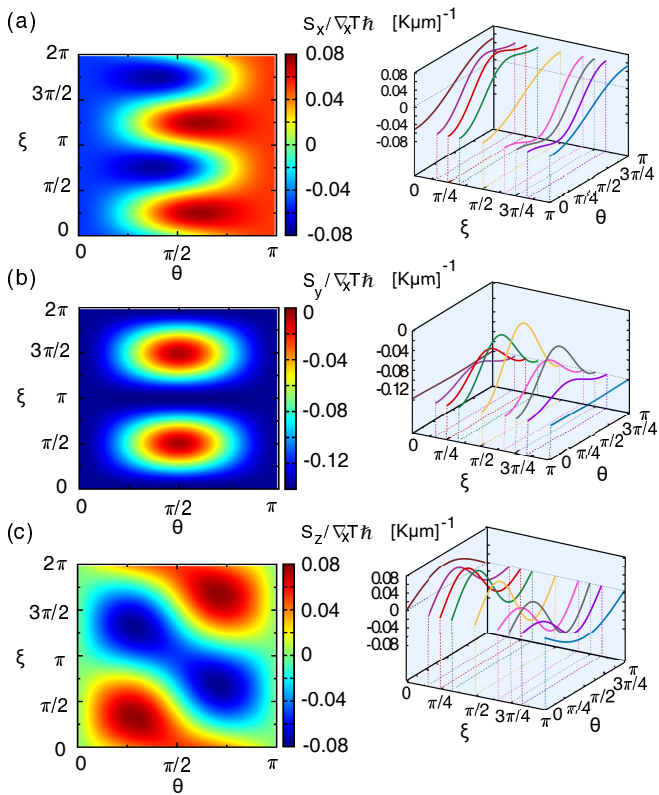}
	\caption{The three components of thermally-induced spin polarization as a function of the polar and azimuthal angles describing orientation of the exchange field. The right panel present some cross-sections of the density plots shown in the left panel. The other parameters are as follows: $m = 0.07 m_{0}$, $\alpha = 8 \cdot 10^{-11}$ eVm, and $\Gamma =10^{-3}$ meV.          }
	\label{fig:fig3}
\end{figure}

In this section we present some general formula for spin polarization.
Inserting Eqs.(\ref{Gf_H})-(\ref{jh_H}) into Eq.(\ref{S_alpha}), taking the trace and integrating over  $\varepsilon$ we obtain the following general formulas for the spin polarization:
\begin{eqnarray}\label{SxT}
S_{x} = \hbar \frac{\nabla_{x} T}{T} \hspace{7cm}\nonumber\\
\times\int \frac{d^{2} \mathbf{k}}{(2\pi)^{2}}\left\{ \frac{1}{2 \Gamma} \left[ \frac{\hbar^{2} k_{x}}{2 m \lambda_{\mathbf{k}}} (\varepsilon_{k} - \mu) + \frac{\alpha}{2 \lambda_{\mathbf{k}}} (\alpha k_{x} - H_{y})\right]\right.\hspace{0.5cm}\nonumber\\
\times (\alpha k_{y} + H_{x}) [f'(E_{+}) - f'(E_{-})]\hspace{0.65cm} \nonumber \\
+\alpha (\varepsilon_{k} - \mu) \left[ \frac{1}{\Gamma} (\alpha k_{x} - H_{y}) (\alpha k_{y} + H_{x}) - H_{z}\right]\hspace{0.5cm}\nonumber\\ \times\frac{f'(E_{+}) + f'(E_{-})}{(2 \lambda_{\mathbf{k}})^{2} + (2 \Gamma)^{2}} \hspace{0.65cm}\nonumber\\
+\frac{1}{2 \Gamma} \frac{\hbar^{2} k_{x}}{2 m} (\alpha k_{y} + H_{x}) [f'(E_{+}) + f'(E_{-})]\hspace{0.7cm} \nonumber\\
\left. + \alpha H_{z} \frac{\varepsilon_{k} - \mu}{4 \lambda_{\mathbf{k}}^{3}} [f(E_{+}) - f(E_{-})]\right\},\hspace{0.84cm}
\end{eqnarray}
\begin{eqnarray}\label{SyT}
S_{y} = \hbar \frac{\nabla_{x} T}{T} \hspace{7cm} \nonumber\\
\times \int \frac{d^{2} \mathbf{k}}{(2\pi)^{2}}\left\{ - \frac{1}{2 \Gamma} \left[ \frac{\hbar^{2} k_{x}}{2 m \lambda_{\mathbf{k}}} (\varepsilon_{k} - \mu) + \frac{\alpha}{2 \lambda_{\mathbf{k}}} (\alpha k_{x} - H_{y}) \right]\right.\hspace{0.3cm} \nonumber\\ \times (\alpha k_{x} - H_{y}) [f'(E_{+}) - f'(E_{-})] \hspace{0.45cm}\nonumber\\
- \frac{1}{2 \Gamma} \left[ \frac{\alpha}{2 \lambda_{\mathbf{k}}^{2}} (\varepsilon_{k} - \mu) (\alpha k_{x} - H_{y}) + \frac{\hbar^{2} k_{x}}{2 m}\right]\hspace{0.35cm}\nonumber\\
\times (\alpha k_{x} - H_{y}) [f'(E_{+}) + f'(E_{-})]\hspace{0.45cm} \nonumber\\
+\frac{\alpha}{2} (\varepsilon_{k} - \mu) \left[\frac{(\alpha k_{x} - H_{y})^{2}}{\lambda_{\mathbf{k}}^{2}} - 1 \right] \hspace{0.35cm}\nonumber\\
\times \left.\frac{2 \Gamma}{ (2 \lambda_{\mathbf{k}})^{2} + (2 \Gamma)^{2}} [f'(E_{+}) + f'(E_{-})]\right\},\hspace{0.8cm}
\end{eqnarray}
\begin{eqnarray}\label{SzT}
S_{z}  = \hbar \frac{\nabla_{x} T}{T}\hspace{7cm}\nonumber\\
\times \int \frac{d^{2} \mathbf{k}}{(2\pi)^{2}}\left\{  \frac{1}{2 \Gamma} \left[ \frac{\hbar^{2} k_{x}}{2 m \lambda_{\mathbf{k}}} (\varepsilon_{k} - \mu) + \frac{\alpha}{2 \lambda_{\mathbf{k}}} (\alpha k_{x} - H_{y})\right]\right. \nonumber\\ \times H_{z} [f'(E_{+}) - f'(E_{-})]\hspace{0.19cm}  \nonumber\\
- \alpha (\varepsilon_{k} - \mu) \left[ (\alpha k_{y} + H_{x}) - \frac{1}{\Gamma} H_{z} (\alpha k_{x} - H_{y})\right]\hspace{0.1cm}\nonumber\\ \times \frac{f'(E_{+}) + f'(E_{-})}{(2 \lambda_{\mathbf{k}})^{2} + (2 \Gamma)^{2}} \hspace{0.25cm}\nonumber\\
+ \frac{1}{2 \Gamma} \frac{\hbar^{2} k_{x}}{2 m} H_{z} [f'(E_{+}) + f'(E_{-})]\hspace{0.31cm} \nonumber\\
\left. - \alpha (\alpha k_{y} + H_{x}) \frac{\varepsilon_{k} - \mu}{4 \lambda_{\mathbf{k}}^{3}} [f(E_{+}) - f(E_{-})]\right\}.\hspace{0.8cm}
\end{eqnarray}
The above general formulas contain information on the behaviour of nonequilibrium spin polarization for arbitrarily oriented exchange field. These formulas will be used later  to calculate numerically the spin polarization for arbitrary orientation of the exchange field.
Thse formulas simplify for some specific orientations of the exchange field. Especially interesting is the situation with the exchange field normal to the plane of 2DEG, so we will restric ourselves to this particular situation.

When the exchange field is oriented perpendicularly to the surface of two-dimensional gas we get:
\begin{eqnarray}
S_{x} =  \alpha \hbar \frac{\nabla_{x}T}{T} H_{z} \int \frac{d^{2} \mathbf{k}}{(2\pi)^{2}} \left\{(\varepsilon_{k} - \mu) \frac{f'(E_{+}) + f'(E_{-})}{(2\zeta)^{2} + (2\Gamma)^{2}} \right.\nonumber\\
+ \left. \frac{\varepsilon_{k} - \mu}{4 \zeta^{3}} [f(E_{+}) - f(E_{-})] \right\},\hspace{0.8cm}
\end{eqnarray}
\begin{eqnarray}
S_{y} = \hbar \frac{\nabla_{x}T}{T} \frac{1}{2\Gamma} \int \frac{d^{2} \mathbf{k}}{(2\pi)^{2}} \left\{- \frac{\hbar}{2 \zeta} \left[ \alpha \frac{\hbar k_{x}^{2}}{m} (\varepsilon_{k} - \mu) + \frac{\alpha}{\hbar} \alpha^{2} k_{x}^{2}\right] \right.\nonumber\\
\times [f'(E_{+}) - f'(E_{-})]
\nonumber\\
- \frac{\hbar}{2 \zeta^{2}} \left[ \frac{\alpha}{\hbar} (\varepsilon_{k} - \mu) \alpha^{2} k_{x}^{2} + \alpha \frac{\hbar k_{x}^{2}}{m} \zeta^{2} \right] [f'(E_{+}) + f'(E_{-})]
\nonumber\\
+\left.  \frac{\alpha}{2} (\varepsilon_{k} - \mu) \left[\frac{\alpha^{2} k_{x}^{2}}{\zeta^{2}} - 1 \right] \frac{f'(E_{+}) + f'(E_{-})}{1 + (\zeta/\Gamma)^{2}} \right\},\hspace{1.0cm}
\end{eqnarray}
\begin{eqnarray}
S_{z} = 0 , \hspace{6.5cm}
\end{eqnarray}
where $\zeta$ is defined as $\sqrt{J^{2} M^{2} + \alpha^{2} k^{2}}$. Now, the component normal to the plane of 2DEG (i.e. along the exchange field) vanishes exactly. The in-plane component normal to the temperature gradient is modified by the exchange field, and additionaly the component along the temperature gradient ($S_x$) appears.

\subsection{Numerical results}

Let us begin with numerical results on spin polarization in case when the exchange field is normal to the plane of 2DEG.
The corresponding results are shown in Fig.~\ref{fig:fig2}. Now, both in-plane components are nonzero. The component $S_y$, which is the only nonvanishing component in the nonmagnetic case, is modified by the exchange field (see the lower panel in Fig.~\ref{fig:fig2}).  This modification is significant especially for chemical potentials inside the energy gap. Such behaviour is in agreement with results obtained recently for oxide perovskites~\cite{Jia2016}.  Additionally, the $S_x$ component appears (see the upper panel in Fig.~\ref{fig:fig2}).

Consider first the component $S_y$. Due to modified electronic spectrum by the exchange field, the spin polarization remarkably depends on $JM$. First, the magnitude of the negative peak of $S_y$ decreases with increasing $JM$. Second, width of the  peaks increases with increasing $JM$. Third, due to a gap of magnitude $2JM$ created by the exchange field at $k=0$, the spin polarization changes sign and is positive in a certain range of positive chemical potentials, see  Fig.~\ref{fig:fig2}b (right of the main negative peak).  These features are also clearly seen in Fig.~\ref{fig:fig2}f,g. The temperature dependence is qualitatively similar to that in the nonmagnetic case, see Fig.~\ref{fig:fig3}d.
In turn, the component $S_x$ is solely due to exchange field and is roughly one order of magnitude smaller than the $S_y$ component, compare upper and lower panels in Fig.~\ref{fig:fig2}.  The dependence of spin polarization on the chemical potential, exchange field and Rashba parameter is qualitatively similar to that of the $S_y$ component, so we will not describe it in more details.

Now we present numerical results on spin polarization for arbitrary orientation of the exchange field. All the three components of spin polarization are shown in Fig.~\ref{fig:fig3}. The right column presents cross-sections of the corresponding density plots in the left panel, which correspond to exchange field oriented in some specific planes. From the density plots one can get the information on the orientation of the exchange field where the spin polarization is maximal. This might be important for description of magnetic dynamics induced by spin torque originated from spin polarization. Such a torque is created owing to exchange coupling of the thermally-induced spin polarization and magnetization. Since the induced spin polarization generally depends on the orientation of exchange field, this torque can be decomposed into field-like and damping-like components.

\section{Summary and conclusions}

We analyzed the spin polarization driven by a temperature gradient in a magnetized 2DEG with Rashba spin-orbit interaction. The limit of a nonmagnetic 2DEG has also been studied in detail. This limit was already studied earlier,\cite{dyrdal13} but some approximations concerning the limit of small Rashba parameter turned out to be not adequate. Therefore, we have reconsidered this limit here in more detail and obtained results which properly describe the temperature dependence of the spin polarization. More specifically, it is shown that the thermal spin polarizability vanishes in the limit of $T=0$.  We considered the impurity vertex corrections to the spin polarization and found that these corrections play an important role.
 	
For a magnetized 2DEG we  calculated the spin polarization for an arbitrary orientation of the exchange field, when all three components of the spin polarization can be nonzero. Such a general situation is important from the point of view of magnetic dynamics. Since the spin polarization leads to a spin torque exerted on the magnetization, the results can be useful when considering magnetic dynamics driven by an external thermal gradient. The torque due to spin polarization can be presented generally as a sum of field-like and damping/antidamping terms -- similarly as in the spin-orbit torques driven by an external electric field or spin transfer torques driven by electric field in spin valves.

We  note that the physical origin of the spin polarization due to a thermal gradient is different from that of the spin polarization driven by an external electric field. In the former case the spin polarization is driven by a statistical force,  while in the latter case this is an electrical force. As numerical  calculations show, the spin polarization induced by a temperature gradient reveals a peak whose maximum is around the band edge of the lower Rashba subband, where asymmetry between the subbands generated by the Rashba coupling is the largest.

\begin{acknowledgments}
	This work was supported by  the Polish Ministry
	of Science and Higher Education through a research
	project Iuventus Plus in years 2015-2017 (project
	No. 0083/IP3/2015/73) and the German Research Foundation (No. SFB 762). 
	A.D. acknowledges the support from the Fundation for Polish Science (FNP). 
	V. D. acknowledges support from the National Science Center 
	in Poland under Grant No. DEC-2012/06/M/ST3/00042.
\end{acknowledgments}


\end{document}